\documentclass[sigconf,preview=true]{acmart}

\renewcommand\footnotetextcopyrightpermission[1]{} 
\AtBeginDocument{%
  }

\setcopyright{none}
\copyrightyear{2025}
\acmYear{2025}

\usepackage{algorithm}
\usepackage{algpseudocode}
\usepackage{listings}
\usepackage{xcolor}

\usepackage{caption}
\usepackage{calc}
\usepackage{multirow}
\usepackage{tabularx}
\usepackage{tikz}
\usetikzlibrary{arrows.meta, positioning, quotes, shapes}

\usepackage{shortcuts}
\usepackage{cleveref}
\crefname{theorem}{Thm.}{Thms.}
\crefname{lemma}{Lem.}{Lemmas}
\crefname{corollary}{Cor.}{Cors.}
\crefname{figure}{Figure}{Figures}
\crefname{definition}{Defn.}{Defns.}
\crefname{table}{Table}{Tables}
\crefname{section}{Section}{Sections}
\crefname{example}{Ex.}{Exs.}
\crefname{item}{item}{items}
\crefname{footnote}{footnote}{footnotes}
\crefname{observation}{Obs.}{Obs.}
\crefname{remark}{Remark}{Remarks}
\crefname{proposition}{Prop.}{Props.}
\crefname{equation}{Eqn.}{Eqns.}
\crefname{counterexample}{Counterexample}{Counterexamples}
\crefname{algorithm}{Algorithm}{Algorithms}
\crefname{principle}{Principle}{Principles}
\crefname{challenge}{Challenge}{Challenges}
\crefname{fact}{Fact}{Facts}

\makeatletter
\newtheoremstyle{acmremark}%
  {.5\baselineskip\@plus.2\baselineskip\@minus.2\baselineskip}
  {.5\baselineskip\@plus.2\baselineskip\@minus.2\baselineskip}
  {\itshape}
  {\z@}
  {\itshape}
  {.}
  {.5em}
  {\thmname{#1}\thmnumber{ #2}\thmnote{ {\normalfont(#3)}}}
\makeatother
\AtEndPreamble{%
  \theoremstyle{acmplain}%
  \theoremstyle{acmdefinition}%
  \theoremstyle{acmremark}%
  \theoremstyle{acmplain}%
}
\AtBeginEnvironment{example}{%
  \pushQED{\qed}%
}
\AtEndEnvironment{example}{\popQED\endexample}

\definecolor{codegreen}{rgb}{0,0.6,0}
\definecolor{codegray}{rgb}{0.5,0.5,0.5}
\definecolor{codepurple}{rgb}{0.58,0,0.82}
\definecolor{backcolour}{rgb}{0.95,0.95,0.92}

\lstdefinestyle{mystyle}{
    backgroundcolor=\color{backcolour},   
    commentstyle=\color{codepurple},
    keywordstyle=\color{magenta},
    numberstyle=\tiny\color{codegray},
    stringstyle=\color{codepurple},
    basicstyle=\ttfamily\footnotesize,
    breakatwhitespace=false,         
    breaklines=true,                 
    captionpos=t,                    
    keepspaces=true,                 
    numbers=left,                    
    numbersep=5pt,                  
    showspaces=false,                
    showstringspaces=false,
    showtabs=false,                  
    tabsize=2
}

\newcommand{\ourmethod}{\textsc{ChatFuMe}} 
\newcommand{\fume}{FUME}
\newcommand{\chatafl}{\textsc{ChatAFL}}

\begin{document}

\title{LLM-Assisted Model-Based Fuzzing of Protocol Implementations}

\author{Changze Huang}
\email{hcz@stu.pku.edu.cn}
\affiliation{%
  \institution{Key Lab of HCST (PKU), MOE;}
  \institution{SCS, Peking University}
  \city{Beijing}
  \country{China}
}

\author{Di Wang}
\email{wangdi95@pku.edu.cn}
\affiliation{%
  \institution{Key Lab of HCST (PKU), MOE;}
  \institution{SCS, Peking University}
  \city{Beijing}
  \country{China}
}

\author{Zhi Quan Zhou}
\email{george.zhou@nio.com}
\affiliation{%
 \institution{NIO Inc.}
 \city{Shanghai}
 \country{China}
 }


\begin{abstract}
Testing network protocol implementations is critical for ensuring the reliability, security, and interoperability of distributed systems. Faults in protocol behavior can lead to vulnerabilities and system failures, especially in real-time and mission-critical applications. A common approach to protocol testing involves constructing Markovian models that capture the state transitions and expected behaviors of the protocol. However, building such models typically requires significant domain expertise and manual effort, making the process time-consuming and difficult to scale across diverse protocols and implementations.

We propose a novel method that leverages large language models (LLMs) to automatically generate sequences for testing network protocol implementations. Our approach begins by defining the full set of possible protocol states, from which the LLM selects a subset to model the target implementation. Using this state-based model, we prompt the LLM to generate code that produces sequences of states. This program serves as a protocol-specific sequences generator. The sequences generator then generates test inputs to call the protocol implementation under various conditions. 
We evaluated our approach on three widely used network protocol implementations and successfully identified 12 previously unknown vulnerabilities. We have reported them to the respective developers for confirmation. This demonstrates the practical effectiveness of our LLM-assisted fuzzing framework in uncovering real-world security issues.
\end{abstract}






\maketitle
\pagestyle{plain}

\section{Introduction}
\label{sec:intro}

Network-protocol implementations are widely deployed across systems from cloud services and web applications~\cite{balakrishnan1995improving, berners1996rfc1945} to embedded devices~\cite{gislason2008zigbee,bluetooth2010bluetooth} and industrial control systems~\cite{thomas2008introduction}.
Bugs in these implementations can persist for an extended period, compromising the security and stability of the systems.
Studies show that even a single malformed input or unhandled edge case can disrupt services and cause catastrophic consequences~\cite{zhang2024survey, gupta2015survey, berger2021survey}.
Thus, identifying these bugs is an essential step when implementing protocols.

In this paper, we focus on \emph{fuzzing} of protocol implementations.
Studies show that fuzzing is an effective method for testing protocol implementations~\cite{pham2020aflnet, fang2021ics3fuzzer, garbelini2022braktooth, garbelini2020greyhound}.
In these methods, a fuzzer usually sends crafted protocol messages over a network interface or directly to a broker to trigger unexpected states, crashes, or abnormal responses.
One major benefit of fuzzing is its capacity to explore a large state space without requiring a formal protocol specification.

Nevertheless, having domain knowledge about protocols can be helpful.
\emph{Model-based fuzzing} of protocol implementations~\cite{natella2022stateafl, amusuo2023systematically, ammann2024dy, pearson2022fume} is a method that uses knowledge about protocols.
These methods use a predefined (usually coarse-grained) model of protocol behavior to generate \emph{sequences} of messages, systematically exploring the state space of the protocol.
One widely employed family of models is finite-state machines (FSMs)~\cite{de2015protocol, fiterau2020analysis, ren2021z, ruge2020frankenstein}.
These methods use an FSM to represent the states and transitions defined by a protocol, enabling the fuzzer to produce sequences of messages that reflect protocol-specific communication patterns.
Following the predefined model, these methods improve input validity, enhance code coverage, and increase the chances of identifying bugs that depend on specific protocol states or message sequences.

However, the reliance on predefined models is a double-edged sword: the applicability and effectiveness of model-based fuzzing depend on the availability and quality of these models.
Adapting model-based fuzzing to different protocols requires the manual construction of protocol models, and adjusting these models to optimize fuzzing performance would be an effort-consuming task.

In this paper, we propose \ourmethod{}, a model-based protocol-implementation fuzzing method that leverages large language models (LLMs)~\cite{brown2020language} to construct and adjust protocol models \emph{automatically}.
LLMs have emerged as a novel tool for general fuzzing~\cite{xia2024fuzz4all, yang2024whitefox}, as well as protocol-implementation fuzzing~\cite{ma2024one, meng2024large,wang2024llmif}.
These LLM-assisted fuzzing methods leverage the understanding and generation capabilities of LLMs to produce syntactically and semantically valid messages for different protocols.
Unlike \ourmethod{}, none of these are model-based: they rely on LLMs to directly generate a large number of sequences of protocol messages, resulting in long generation time and high token consumption.

Instead of direct generation of message sequences, \ourmethod{} uses LLMs to automatically construct and adjust protocol models in the form of random sequence-generator \emph{programs} that generate random message sequences.
Our motivation is driven by the LLMs' strong capabilities in generating \emph{programs}, inferring state transitions, and embedding domain knowledge from protocol documentation and usage patterns.
Our design of \ourmethod{} aims to strike a balance among the following desiderata:
\begin{itemize}
  \item \emph{Flexibility}. \ourmethod{} handles implementations of different protocols with little adaptation effort.
  \item \emph{Effectiveness}. \ourmethod{} achieves comparable performance against model-based fuzzing with predefined models.
  \item \emph{Cost-efficiency}. \ourmethod{} consumes much fewer tokens than prior LLM-assisted protocol fuzzing methods.
\end{itemize}

\ourmethod{} consists of two major components: (i) automatic model construction, and (ii) feedback-guided fuzzing loop.
The model construction starts with identifying protocol states, using the protocol's documentation (possibly with some user prompts to encode domain knowledge) as input.
\ourmethod{} then asks the LLM to summarize the protocol behavior, capturing high-level domain-specific patterns of the states and transitions.
Next, \ourmethod{} prompts the LLM to select key states and summarize the transition rules among them.
These states and transitions form the basis of a coarse-grained protocol model.
Rather than directly generating message sequences, \ourmethod{} asks the LLM to construct a sequence generator, which is an executable program that samples state transitions and generates random message sequences.

After the model construction, \ourmethod{}'s fuzzing loop starts with pairing the random sequence generator with a user-provided payload generator that handles the low-level field formatting of messages.
The user can once again use an LLM to program the payload generator in advance.
Executing the random generator multiple times, \ourmethod{} generates multiple message sequences and sends them to the protocol implementation being tested.
\ourmethod{} then collects the behavior of the protocol implementation, such as its responses, and prompts the LLM to evaluate these results and suggest adjustments to the protocol model.
In particular, the LLM is supposed to provide feedback on whether to add new states, remove existing ones, or update transition probabilities.
With the feedback, \ourmethod{}'s model-construction component adjusts the protocol model and the generator program.
The fuzzing-loop component then begins another iteration using the adjusted model.

Our experiments demonstrate that \ourmethod{} is reasonably flexible, effective, and cost-efficient.
In testing three different real-world protocol implementations, our approach discovered 12 potential bugs, validating its fault detection capability.
Compared to a prior model-based fuzzer, our method identified more new protocol states within the same time window, demonstrating its effectiveness in exploration.
Additionally, compared to a prior LLM-based fuzzer, our technique achieves lower token consumption, highlighting its cost efficiency and scalability in practical fuzzing scenarios.

\paragraph{Contributions}
The paper's contributions include the following:
\begin{itemize}
\item We propose \ourmethod{}, an LLM-assisted model-based fuzzing method for protocol implementations.
The key innovation is that it uses LLMs to construct and adjust a protocol-model program that generates message sequences.

\item We implement \ourmethod{} and conduct an experimental evaluation of it. Our experiments demonstrate the flexibility of \ourmethod{} by applying it to three different protocols, its effectiveness by comparing it against a prior model-based method \fume{}~\cite{pearson2022fume}, and its cost efficiency by comparing it against a prior LLM-assisted method \chatafl{}~\cite{meng2024large}.

\item We apply \ourmethod{} on three real-world protocol implementations (HMQ, PyModbus, and Moquette) and discover 12 potential bugs in these implementations.

\end{itemize}
\section{Background: Model-Based Fuzzing of Protocol Implementations}
\label{sec:background}

In this section, we review \fume{}~\cite{pearson2022fume}, a model-based fuzzing technique designed for Message Queuing Telemetry Transport (MQTT) implementations.
We begin by reviewing the MQTT protocol, which we will use as a concrete protocol to demonstrate our method in \cref{sec:method}.
We then sketch \fume{}'s predefined model that guides the fuzzing of MQTT implementations.

\subsection{The MQTT Protocol}
\label{sec:mqtt}

The Message Queuing Telemetry Transport (MQTT) protocol~\cite{mqttMQTTSpecification} has emerged as the de facto standard for messaging in the Internet of Things (IoT) and Industrial IoT (IIoT) domains~\cite{soni2017survey}.
Standardized by OASIS and ISO, MQTT is a lightweight, event-driven protocol designed for environments with limited bandwidth and high latency, making it ideal for devices such as sensors, embedded systems, and industrial PLCs.
At its core, MQTT operates on a publish/subscribe model, which inherently decouples message \emph{publishers} (i.e., senders) from \emph{subscribers} (i.e., receivers).
Communication is facilitated through \emph{topics}, which serve as virtual channels for message exchange.
The central component of this architecture is the MQTT \emph{broker} (i.e., server), which manages connections, filters incoming messages from publishers based on their topics, and efficiently distributes them to all interested subscribers.

In MQTT, all communication between clients and brokers is facilitated through the exchange of \emph{control packets}, which are the fundamental units of data transfer.
These packets encapsulate various operational commands, enabling functions such as establishing connections, managing subscriptions, and publishing application messages.
Each control packet adheres to a structured format consisting of up to three main components: a fixed header (FH), a variable header (VH), and the \emph{payload}, which contains the actual data or message that may vary depending on the control packet type.
MQTT supports 15 different packet types, including connection management, message publishing, and subscription management.
\cref{tab:mqtt-packet-types} summarizes the 15 MQTT packet types.

\begin{table}[t]
\small
\caption{A summary of MQTT control packets.}
\label{tab:mqtt-packet-types}
\begin{tabularx}{\columnwidth}{|@{\hspace{2pt}}p{0.2\columnwidth}@{\hspace{2pt}}|@{\hspace{2pt}}p{0.52\columnwidth}@{\hspace{2pt}}|@{\hspace{2pt}}X@{\hspace{2pt}}|}
  \hline
  \textbf{Name} & \textbf{Purpose} & \textbf{Components} \\ \hline
  CONNECT & Initiate connection to broker & FH, VH, payload \\
  CONNACK & Acknowledge connection request & FH, VH \\
  PINGREQ & Check if connection to broker is alive & FH \\
  PINGRESP & Confirm connection is active & FH \\
  DISCONNECT & Close network connection gracefully & FH \\
  AUTH & Exchange authentication data & FH, VH \\ \hline
  PUBLISH & Deliver message to subscribers & FH, VH, payload \\
  PUBACK & Acknowledge QoS 1 PUBLISH & FH, VH \\
  PUBREC & Acknowledge QoS 2 PUBLISH & FH, VH \\
  PUBREL & Confirm receipt of PUBREC & FH, VH \\
  PUBCOMP & Confirm receipt of PUBREL & FH, VH \\ \hline
  SUBSCRIBE & Request subscription to topics & FH, VH, payload \\
  SUBACK & Acknowledge SUBSCRIBE packet & FH, VH, payload \\
  UNSUBSCRIBE & Request to cancel subscriptions & FH, VH, payload \\
  UNSUBACK & Acknowledge UNSUBSCRIBE packet & FH, VH, payload \\ \hline
\end{tabularx}
\end{table}

The explicit connection management packets highlight MQTT's \emph{stateful} nature.
Such statefulness enables the broker to retain a client's subscriptions and buffer messages for delivery upon reconnection, which is crucial for maintaining persistent sessions.
However, the statefulness results in a vast and complex input space for fuzzing.
A fuzzing method needs domain knowledge about MQTT to generate valid message sequences (e.g., they should start with a CONNECT message) and explore deep states (e.g., some states can only be reached by subscribing to a specific topic).

\subsection{\fume{}'s Fuzzing Model for MQTT Brokers}
\label{sec:fume}

\fume{} is a fuzzing method designed explicitly for testing MQTT implementations~\cite{pearson2022fume}.
To this end, \fume{} is a model-based fuzzing method because it requires a predefined model of the MQTT protocol to guide the fuzzing process.
\fume{}'s model is coarse-grained: it only knows the 15 message types shown in \cref{tab:mqtt-packet-types}, and every message sequence must start with CONNECT.

\fume{} then uses the MQTT model to construct two Markov models (i.e., FSMs with probabilistic transitions) to describe mutation-guided and generation-guided fuzzing, respectively.
\emph{Mutation}-guided fuzzing requires an input corpus of valid message sequences as test cases.
On the other hand, \emph{generation}-guided fuzzing requires domain knowledge of the protocol to generate valid message sequences.
In this paper, we aim to develop a \emph{flexible} fuzzing method applicable to different protocol implementations, where informal protocol descriptions are often more readily available than a corpus of test cases; thus, we focus on generation-guided fuzzing.

\begin{figure}[t]
\includegraphics[width=0.9\columnwidth]{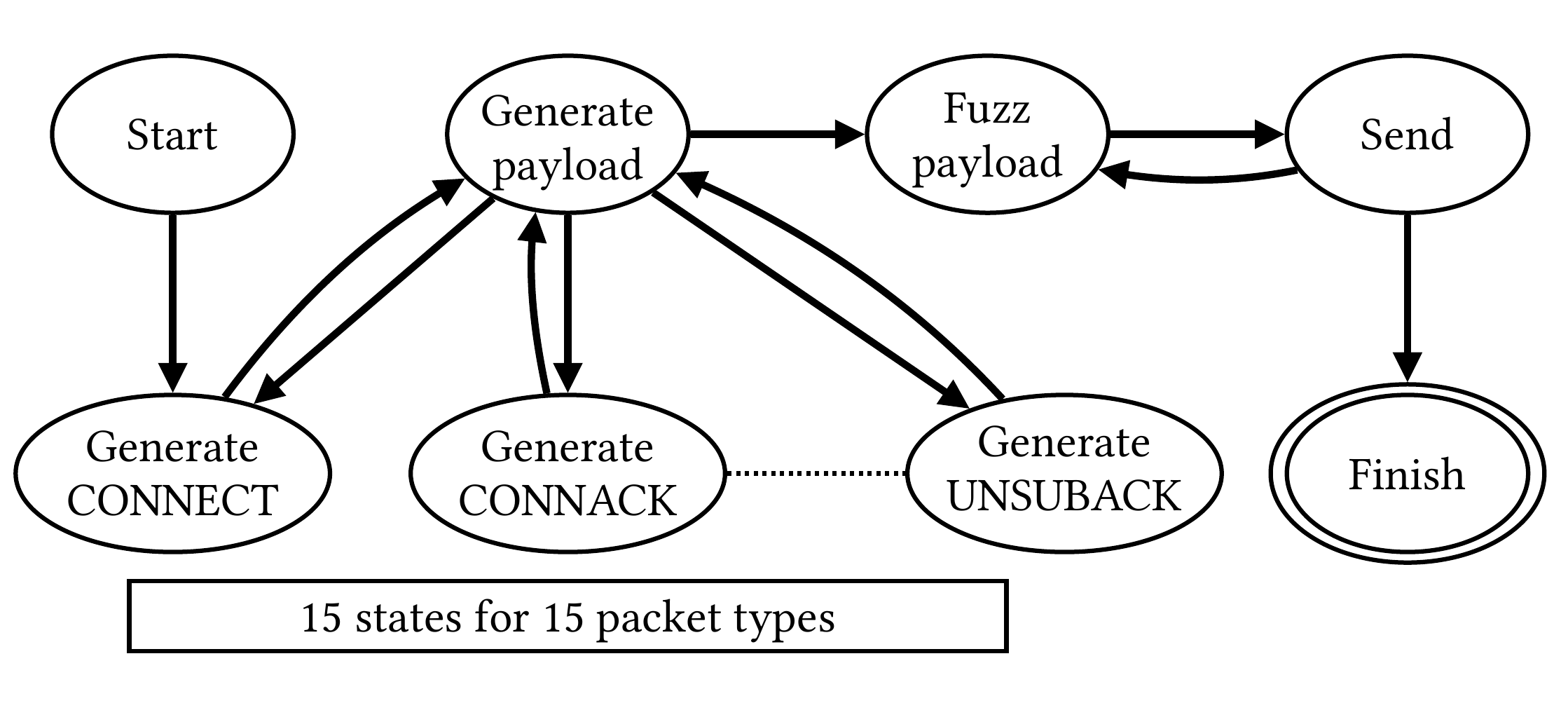}
\caption{\fume{}'s generation-guided fuzzing model.}
\label{fig:fume-model}
\end{figure}

\cref{fig:fume-model} demonstates \fume{}'s fuzzing model of one iteration of generation-guided fuzzing.
We ignore the state-transition probabilities, so it appears just like an FSM.
The model describes a random generation process: it starts with generating a CONNECT message and its payload, then stochastically generates some follow-up messages, applies fuzzing (e.g., insertion, deletion, and mutation) to the payloads multiple times, and finally sends the message sequences to the MQTT broker.
\fume{} has another fuzzing model for its mutation-guided fuzzing process, and it alternates between the two fuzzing models to leverage the strengths of both models.

We take inspiration from \fume{} with a key observation: such fuzzing models can be easily expressed by \emph{executable programs} that generate random message sequences.
Moreover, payload generators are also executable programs.
The observation motivates us to leverage the understanding and programming capabilities of LLMs to extract domain knowledge about protocols and generate executable programs that represent the fuzzing models.
\section{Our Method}
\label{sec:method}

\begin{figure*}[t]
    \centering
    \includegraphics[keepaspectratio=true,width=0.9\textwidth]{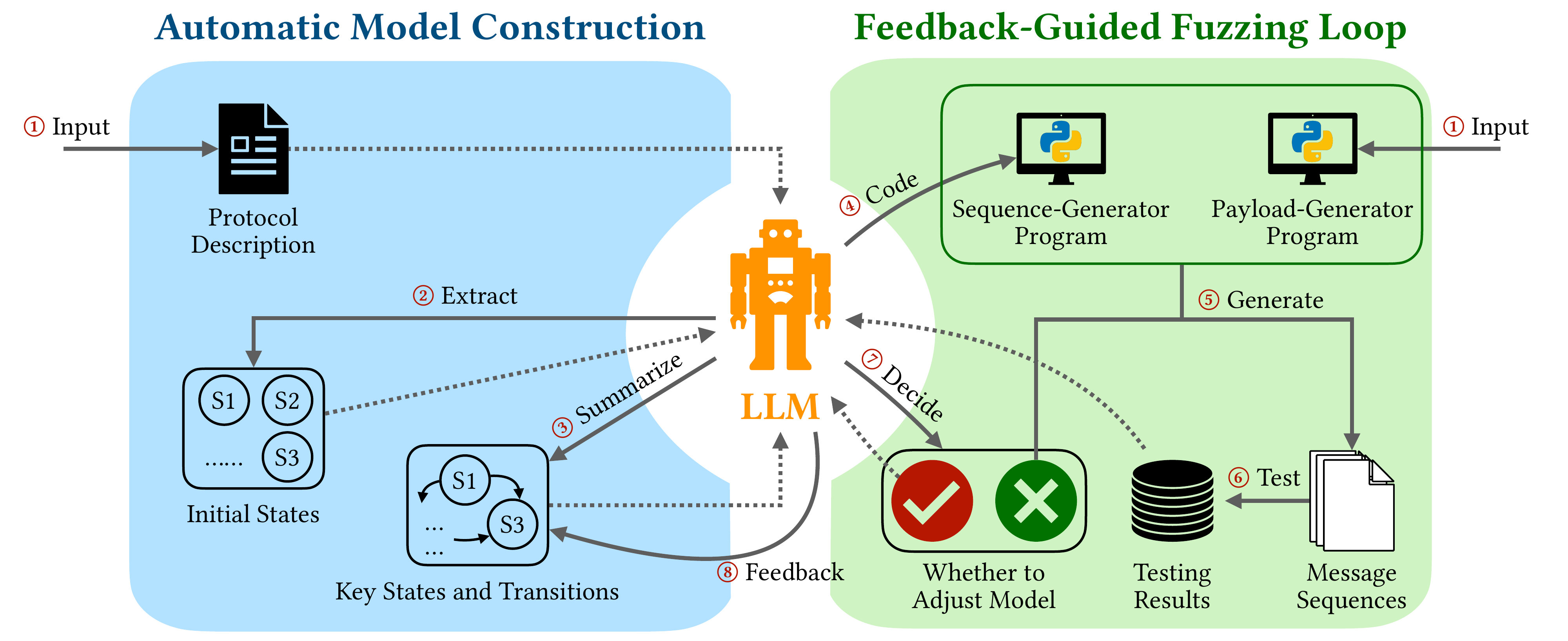}
    \caption{An overview of \ourmethod{}'s workflow. Dotted arrows indicate prompting the LLM for specific tasks.}
    \label{fig:overview}
\end{figure*}

In this section, we describe the workflow and technical details of our \ourmethod{} method.
\cref{sec:workflow} presents an overview of \ourmethod{}'s workflow and its two major components: (i) automatic model construction, and (ii) feedback-guided fuzzing loop.
\cref{amc,afl} use MQTT as a demonstration protocol to explain the two components, respectively.

\subsection{Overview of the Workflow}
\label{sec:workflow}

Figure \ref{fig:overview} presents an overview of \ourmethod{}.
Its key feature is combining model-based protocol-implementation fuzzing with LLMs: it uses LLMs to automate various steps in the workflow, including the construction and adjustment of the fuzzing model.
In this way, \ourmethod{} achieves flexibility to handle different protocols.

\paragraph{Automatic model construction}
\ourmethod{} begins with user-provided protocol documentation, which may include prompts to describe protocol states as a high-level summary of a protocol's specification.
\ourmethod{} uses an LLM to augment the set of states by extracting domain knowledge from the protocol documentation.
With the augmented set of states, \ourmethod{} again requests the LLM to perform state selection and summarize the state-transition rules, narrowing down to a small but essential subset of states as the starting point for fuzzing and capturing how the protocol allows moving between different states.
These transitions, with the selected states, form a lightweight coarse-grained model of protocol behavior.
This component also provides the functionality to adjust the model by prompting an LLM to modify the selected states based on feedback from the fuzzing loop, which we explain below.

\paragraph{Feedback-guided fuzzing loop}
Based on the constructed protocol model, \ourmethod{} uses an LLM to generate a sequence-generator program, which produces randomized state sequences that conform to transition rules.
The sequence generator is paired with a user-provided payload generator---which an LLM could generate in advance---to generate complete message sequences. 
\ourmethod{} then sends these sequences to the protocol implementation and monitors its behavior, e.g., whether it crashes or reports abnormal responses.
After collecting the testing results, \ourmethod{} decides whether to adjust the protocol model.
We employ two mechanisms for this decision:
(i) with a predefined probability, \ourmethod{} resets the protocol model to the initial one constructed at the beginning of the workflow, i.e., incorporates \emph{random restart}; or
(ii) \ourmethod{} prompts the LLM to analyze the testing results and decide if the model needs refinement, and if so, further prompts it to identify states that contribute to effective testing outcomes.
Typically, the LLM would add a few new states or remove existing ones from the protocol model.
If the model gets adjusted, \ourmethod{} loops back to the phase of generating a sequence-generator program via an LLM.
Otherwise, it keeps the sequence-generator program unchanged and starts the next loop iteration.

\paragraph{Incorporation of LLMs}
We use LLMs because they provide capabilities for understanding documentation (for model construction), generating code (for program generation), and analyzing testing results (for model adjustment).
Furthermore, for widely used protocols, even if the user does not possess domain knowledge, LLMs' familiarity with these protocols enables \ourmethod{} to perform fuzzing effectively with minimal guidance.
Instead of asking the LLM to generate individual test cases directly, we prompt it to generate an executable program as a random sequence generator.
This generator-based design significantly reduces token consumption by generating numerous message sequences from a single generator, making \ourmethod{} cost-efficient for LLM-assisted fuzzing of real-world protocol implementations.

\subsection{Automatic Model Construction}
\label{amc}

The construction of our fuzzing model begins with user-provided identification of protocol states, which can be derived from existing work, specifications, documentation, or manual analysis.
Listing \ref{lst:state} provides an example of the initial states of MQTT.
These initial states can be extracted from the MQTT specification~\cite{mqttMQTTSpecification}.
These states provide a coarse-grained foundation for understanding the protocol's high-level behavior.
To enrich this initial model, we prompt the LLM to incorporate domain knowledge, such as typical client-server interactions and expected message sequences.
This step enhances the completeness and semantic accuracy of state definitions, enabling better downstream reasoning during fuzzing.
By combining user input with LLM-driven knowledge expansion, we strike a balance between manual guidance and automated insight.

\begin{lstlisting}[label=lst:state, language=Python, caption=An example of initial states. The list above represents the primary MQTT control packet types. Each state corresponds to a specific control packet used in the communication between MQTT clients and brokers.]
states = [ "CONNECT", "CONNACK", "PUBLISH", 
           "PUBACK", "PUBREC", "PUBREL", 
           "PUBCOMP", "SUBSCRIBE", "SUBACK", 
           "UNSUBSCRIBE", "UNSUBACK", "PINGREQ", 
           "PINGRESP", "DISCONNECT", "AUTH" ]
\end{lstlisting}

Once domain knowledge is incorporated, we leverage the LLM to refine the state space by selecting a concise set of essential protocol states for testing.
This step abstracts the protocol into a manageable number of representative states that retain sufficient semantic coverage while reducing unnecessary complexity.
As illustrated in Listings~\ref{lst:select} and~\ref{lst:example}, we construct a two-part prompt to guide the LLM in identifying essential protocol states for testing. The Listings~\ref{lst:select} provides contextual information about the protocol and the goal of reducing the input space by selecting a small number of high-impact states. The Listings~\ref{lst:example} specifies a structured output format, requiring the LLM to return a JSON array of state names and concise justifications for their importance.
This automated abstraction enables \ourmethod{} to focus on high-impact areas of the protocol, balancing coverage and efficiency in the input space exploration.

\begin{lstlisting}[label=lst:select, language=Python, caption=The prompt for state selection.]
Prompt_for_States_Selection = f'''I am designing a model-based testing framework for the {protocol}. To reduce the search space, I want to abstract the protocol into a small number of essential states that capture the most important aspects of its behavior for testing purposes.
Please help me identify {number} essential states above that cover the core functionality of {protocol} while preserving enough semantics to be useful for testing client and broker implementations. 
Output Format: {example}'''
\end{lstlisting}

\begin{lstlisting}[label=lst:example, language=Python, caption=The Example for state selection.]
Example_for_States_Selection = f'''
Return a JSON array of {number} objects. 
Each object should have ONLY the following fields:
`"select"`: A short name for the state 
(e.g., `"CONNECT"`, `"PUBLISH"`)
`"reason"`: A concise explanation of why this state is 
essential for {protocol} testing
'''
\end{lstlisting}

\subsection{Feedback-Guided Fuzzing Loop}
\label{afl}

In the feedback-guided fuzzing loop, we leverage the LLM to analyze execution results and guide the adjustment of the protocol model.
The loop begins with the LLM-generated sequence generator, which produces message sequences following the current state transition model.
These sequences serve as high-level plans for how a client might interact with a protocol implementation.

To guide the LLM in producing a realistic sequence generator, we apply an \emph{autoprompting} strategy~\cite{brown2020language, wang2022self, xia2024fuzz4all} to create a high-quality prompt that encapsulates the protocol specification.
Given a set of selected states, we ask the LLM to construct a prompt that will later be used to generate Python code implementing a random state sequence generator.
This prompt includes constraints to preserve realistic protocol behavior, such as capturing state transition probabilities, enforcing randomness, and ensuring variable-length sequences.
In this way, we distill the protocol specification into a reusable instruction, forming a bridge between abstract state modeling and a concrete generation of the generator.

After generating a prompt tailored for protocol behavior, we use the LLM to produce a Python program as the sequence generator, which we write into a module and dynamically import for execution.
This integration step serves both as a code validity check and as the mechanism to link LLM output with our fuzzing pipeline.
If the generated code passes import and runtime validation, we invoke it repeatedly in the loop.
Listing \ref{lst:code} presents an example of the code generated by the LLM, which incorporates both the protocol specification and random control structures. 

\begin{lstlisting}[label=lst:code, language=Python, caption=The Python code of an LLM-generated random sequence generator for the MQTT protocol.]
import random

def MQTT_state_generator():
    states = ['CONNECT', 'CONNACK', 'PUBLISH', 'SUBSCRIBE', 'DISCONNECT', 'PINGREQ', 'PUBACK']
    state_sequence = []
    current_state = 'CONNECT'
    state_sequence.append(current_state)

    while True:
        if current_state == 'CONNECT':
            next_state = random.choices(['CONNACK', 'SUBSCRIBE'], weights=[70, 30])[0]
        elif current_state == 'CONNACK':
            next_state = random.choices(['PUBLISH', 'SUBSCRIBE', 'DISCONNECT'], weights=[50, 30, 20])[0]
        elif current_state == 'PUBLISH':
            next_state = random.choices(['PUBACK', 'SUBSCRIBE', 'DISCONNECT'], weights=[60, 30, 10])[0]
        elif current_state == 'SUBSCRIBE':
            next_state = random.choices(['PUBLISH', 'DISCONNECT', 'PINGREQ'], weights=[40, 30, 30])[0]
        elif current_state == 'DISCONNECT':
            if random.random() < 0.5:
                break
            else:
                next_state = 'CONNECT'
        elif current_state == 'PINGREQ':
            next_state = random.choices(['DISCONNECT', 'PUBACK'], weights=[70, 30])[0]
        elif current_state == 'PUBACK':
            next_state = random.choices(['PUBLISH', 'DISCONNECT'], weights=[60, 40])[0]

        state_sequence.append(next_state)
        current_state = next_state

        if random.random() < 0.5:
            break

    return state_sequence
\end{lstlisting}

The generated sequences are then passed to a user-provided payload generator, which translates each sequence into concrete protocol messages.
Note that the payload generator could also be generated in advance by an LLM.
The resulting payloads are sent to the target protocol implementation, and \ourmethod{} records the feedback from each run to inform further model adjustment.

 After sending a sufficient number of fuzzing sequences (ranging from 20,000 to 50,000 in our experiments), \ourmethod{} analyzes the feedback from the target protocol implementation and evaluates the fuzzing effectiveness, guiding model adjustment.
 The analysis focuses on identifying failure patterns by computing statistics such as the number of failures per protocol function, total request distribution, and failure rates.
 Specifically, we categorize responses like timeouts or connection resets as failures and calculate their frequency relative to the total number of generated sequences.
 This analysis provides insight into which parts of the protocol are more error-prone, helping to inform the LLM-driven model adjustment in subsequent fuzzing iterations.
 
The analysis above is then formatted and sent as part of a prompt to the LLM, asking it to interpret the results in the context of the target protocol and suggest whether any states should be added or removed from the model to improve test effectiveness, as shown in Listing \ref{lst:deci}.
Notably, we only ask the LLM to interpret the statistical results and enhance test effectiveness.
We do not explicitly ask it to analyze or reason about correlations between different types of failures.
To ensure reliability, we validate any LLM-suggested states to avoid hallucinated or irrelevant protocol behavior.
This LLM-driven evaluation allows our fuzzing process to adapt intelligently over time, refining the model based on concrete feedback from protocol implementations.

\begin{lstlisting}[label=lst:deci, language=Python, caption=The prompt for model adjustment.]
Prompt_for_Decision = f'''{reuslt_summary}
Above is the summary of fuzzing results of a {protocol} implementation using these states: 
{states}
Do you think I should add or remove more states in the search space? 
Give your result in JSON format. 
It should ONLY have two fields:
`"decision"`: ADD or DELETE answer for should I add more states in search space or delete one state in the search space?
`"reason"`: A concise explanation of why I should add more states.'''
\end{lstlisting}

To enhance exploration and prevent convergence to a local optimum, we introduce controlled randomness into the loop by occasionally re-initializing the model from the beginning, allowing the system to escape stagnant state configurations. (Note that this mechanism is not illustrated in \cref{fig:overview}.)
The fuzzing loop continues until a crash or critical fault is observed, ensuring prolonged exploration when necessary.
When analyzing results, if adjustment is needed, we prompt the LLM to suggest a new state to add (from previously unselected candidates) or recommend removing underperforming states.
These decisions are based on summarized feedback from past executions, and each recommendation includes a justification to preserve model clarity.
By integrating randomness and iterative adjustment, the loop strikes a balance between exploiting known effective sequences and exploring new protocol behaviors.

\section{Experimental Design}

In this section, we describe our experimental design to evaluate our \ourmethod{} method.
We propose the following three research questions, concerning flexibility, effectiveness, and cost efficiency:

\begin{itemize}
\item \textbf{RQ1}: How flexible and effective is \ourmethod{} in discovering faults across different protocols and implementations?

\item \textbf{RQ2}: How does \ourmethod{} compare to a prior model-based fuzzing method in protocol testing effectiveness?

\item \textbf{RQ3}: What are the characteristics of \ourmethod{} in terms of token usage compared with an existing LLM-based fuzzer? 

\end{itemize}

\subsection{Systems Under Test and Baselines}

We selected three network protocols and six real-world software implementations in total, with varying levels of maturity and popularity, as reflected by their GitHub star counts. Table~\ref{tab:suts} presents the statistics, including the protocol and programming language they are based on, their popularity (GitHub stars), and the specific RQs they were used to evaluate. This setup helps ensure that \ourmethod{} is not tied to any single protocol or implementation style, reinforcing its flexibility and practical applicability.

\begin{table}[t]

\caption{A summary of protocol implementations used in our evaluation. ``\#Stars'' indicates the number of GitHub stars (as of July 2025). ``Used in'' denotes which research question(s) (RQ1--RQ3) each implementation contributed to.}
\label{tab:suts}

\begin{tabular}{l|llll}
\hline
\textbf{Name}      & \textbf{Protocol} & \textbf{Language}   & \textbf{\#Stars} & \textbf{Used in}    \\ \hline
HMQ       & MQTT     & Go         & 1359    & RQ1        \\
Moquette  & MQTT     & Java       & 2382    & RQ1        \\
Mosquitto & MQTT     & C          & 9928    & RQ1 \& RQ2 \\
Aedes     & MQTT     & JavaScript & 1873    & RQ2        \\
Pymodbus  & Modbus   & Python     & 2494    & RQ1        \\
OwnTone   & DAAP     & C          & 2287    & RQ1 \& RQ3 \\ \hline
\end{tabular}

\end{table}

\paragraph{RQ1}
To demonstrate the flexibility and effectiveness of our approach, we conduct experiments across three different protocol implementations: MQTT, Modbus, and Digital Audio Access Protocol (DAAP). These protocols span different formats, transport layers, and usage domains. The goal of this part of the experiment is to demonstrate that \ourmethod{} is not limited to any specific protocol or domain. 

MQTT is a lightweight publish-subscribe messaging protocol commonly used in IoT systems.
We evaluated our approach on three different MQTT broker implementations: HMQ~\cite{joy2025fhmq}, Moquette~\cite{moquetteioMoquetteBroker}, and Mosquitto~\cite{mosquittoEclipseMosquitto}.
This demonstrates \ourmethod{}'s flexibility across languages and ecosystems.
HMQ is a high-performance MQTT broker written in Go, designed for scalability and compatibility with MQTT 3.1.1 and standard clients.
Moquette is a lightweight, embeddable Java broker that supports MQTT versions 3 and 5, featuring session expiration and topic aliasing.
Finally, Mosquitto is a widely used C implementation that offers a compact MQTT broker and client suite.
Testing across these diverse implementations helps establish that \ourmethod{} is protocol-agnostic and language-agnostic.
It is capable of handling varying runtime environments and codebases.

Modbus is an industrial control protocol based on function codes.
We selected Pymodbus~\cite{githubGitHubPymodbusdevpymodbus} as the target implementation for Modbus in our evaluation.
Pymodbus is a full-featured, open-source Modbus protocol stack written in Python, supporting both synchronous and asynchronous APIs.
It provides built-in client and server simulators, payload builder/decoder functions, and supports both standard and extended Modbus function codes with minimal external dependencies. 

DAAP is a binary protocol layered over HTTP used for media sharing.
We chose OwnTone~\cite{owntoneOwnTone} as our DAAP server for evaluation.
OwnTone is an open-source media server written in C, designed to serve audio content over the DAAP.
It supports sharing and streaming music via DAAP, making it a versatile and realistic target for fuzz testing.
By incorporating OwnTone as our test subject, we demonstrate that our framework can handle binary protocol implementations layered over HTTP.

\paragraph{RQ2}
To evaluate the testing effectiveness of our approach, we compare it against \fume{}, a model-based fuzzing technique explicitly designed for MQTT.
\fume{} combines mutation-based and generation-based fuzzing strategies and introduces Markov chains to guide both payload mutation and generation. It models the fuzzing process as a finite Bernoulli process to explore MQTT protocol behaviors and uncover vulnerabilities thoroughly.

This research question is evaluated in two parts.
First, we measure and compare the number of test cases generated by each method within a fixed time window, assessing the throughput and exploration capability of the fuzzers. Second, we analyze the crash discovery speed, which is the rate at which each tool triggers a fault or crash in the target protocol implementation.
For the throughput comparison, we evaluate on Mosquitto. For the crash speed comparison, we use Aedes~\cite{githubGitHubMoscajsaedes}, a popular MQTT broker implemented in JavaScript.

\paragraph{RQ3}
For token usage analysis, we compare our approach with \chatafl{}~\cite{meng2024large}, a recent LLM-guided protocol fuzzer.
\chatafl{} leverages large language models trained on human-readable protocol specifications to extract protocol message grammars and predict stateful interactions.
It uses LLMs to generate message sequences and detect states in protocol implementations, combining grammar construction with mutation and sequence prediction.
We conduct a comparison focused on: the total number of tokens consumed and the number of LLM API calls required during the fuzzing process.

\subsection{Our Implementation}
\ourmethod{} is primarily implemented in Python, with an emphasis on cost-efficiency and broad applicability.
To maximize flexibility, all experiments in RQ1 and RQ2 were conducted using GPT-4o-mini, a lightweight language model that is readily interchangeable with many popular alternatives on the market.
This is made possible by our design choice to decompose the overall fuzzing workflow into smaller, modular tasks.
This eliminates the need for large token windows or high-capacity models.
For RQ3, we additionally evaluated \ourmethod{} using GPT-3.5 Turbo, which successfully handled all required subtasks, further demonstrating the adaptability of our approach to different LLM configurations.
In all experiments, the temperature parameter was set to 0.5 to introduce controlled randomness and encourage diverse outputs from the LLM.

We obtain the initial states and payload generators for MQTT, Modbus, and Digital Audio Access Protocol (DAAP) using the following approach:
\begin{itemize}
\item For MQTT, we extracted protocol states directly from the official specification\cite{mqttMQTTSpecification} and adopted the existing payload generator from \fume{}.
\item For Modbus, we similarly derived states from the official specification \cite{modbusModbusSpecifications}. To generate payloads, we provided ChatGPT with examples from the specification. ChatGPT generates a protocol-aware payload generator totaling 501 lines of code.
\item For DAAP, which is layered over HTTP and features loosely structured binary payloads, we used ChatGPT to create a lightweight generator based on the specification\cite{githubGitHubBjoernricksdaapprotocol} and integrated simple mutation strategies. The resulting generator is compact---just over 100 lines of code---yet effective.
\end{itemize}

\subsection{Experimental Setup}
We designed fuzzing campaigns tailored to each research question while balancing resource constraints and consistency. 

\paragraph{RQ1} We ran each fuzzing campaign for five hours to allow sufficient exploration while limiting the cost of LLM API calls due to hardware and budget constraints. 

\paragraph{RQ2} We conducted a more controlled comparison with \fume{}. To mitigate randomness, we ran both \ourmethod{} and \fume{} for one hour, repeated across three independent trials, and measured the total number of unique test cases, the total number of test cases generated, the average test case length, and the new response found. To evaluate the crash discovery speed, we ran both tools three times and recorded the time it took for each to trigger the first crash. 

\paragraph{RQ3} Since \chatafl{} does not store all generated test cases, we cannot directly compare the number of test cases over a fixed time. Therefore, we focus on comparing token usage and the number of LLM calls. To ensure a fair comparison, we use GPT-3.5 Turbo instead of GPT-4o-mini in our evaluation.

\paragraph{Environment} All experiments were conducted on a virtual machine running Ubuntu 20.04. The VM was allocated 11.4 GB of memory, 4 CPU cores, and a 50 GB SCSI hard disk. The host machine is equipped with a 13th Gen Intel(R) Core(TM) i9-13900H @ 2.60 GHz and 32 GB of RAM. All LLM calls in our experiments were made via the OpenAI API.

\paragraph{Metrics} Because \ourmethod{} does not rely on instrumentation, we do not report traditional code coverage metrics~\cite{wei2022free, bohme2022reliability}. For the comparison in RQ2 regarding test case generation, we measure the total number of test cases generated, the number of unique test cases, the average test case length, and the number of distinct responses discovered. For RQ3, we focus on efficiency metrics by comparing the total number of LLM tokens used and the number of LLM API calls made.

\begin{table*}[thb]
\centering

\caption{A summary of identified potential bugs by protocol, software, error type, and description.}
\label{tab:bug_summary}

\begin{tabular}{c|l|l|l|l}
\hline
\textbf{\#} & \textbf{Protocol} & \textbf{Subject} & \textbf{Error Type} & \textbf{Description} \\ \hline
1  & Modbus & PyModbus & \texttt{struct.error} & Buffer too small for \texttt{>HH} in \texttt{register\_message.py:180}. \\ 
2  & Modbus & PyModbus & \texttt{struct.error} & Buffer too short for \texttt{>HHB} in \texttt{bit\_message.py:134}. \\ 
3  & Modbus & PyModbus & \texttt{struct.error} & Buffer too short for \texttt{>H} in \texttt{file\_message.py:238}. \\ 
4  & Modbus & PyModbus & \texttt{struct.error} & Buffer too short for \texttt{>BBB} in \texttt{mei\_message.py:52}. \\ 
5  & Modbus & PyModbus & \texttt{struct.error} & Buffer too short for \texttt{>HH} in \texttt{bit\_message.py:30}. \\ 
6  & Modbus & PyModbus & \texttt{struct.error} & Buffer too short \texttt{>HH} in \texttt{register\_message.py:26}. \\ 
7  & Modbus & PyModbus & \texttt{struct.error} & Buffer too short for \texttt{>BHHH} in \texttt{file\_message.py:61}. \\ 
8  & Modbus & PyModbus & \texttt{struct.error} & Buffer too short for \texttt{>HHB} in \texttt{register\_message.py:225}. \\ 
9  & MQTT   & HMP      & \texttt{nil pointer dereference} & Crash if \texttt{conn.RemoteAddr()} is nil  \\
10 & MQTT   & Moquette  & \texttt{IOException} & Invalid MQTT message caused channel closure. \\ 
11 & MQTT   & Moquette  & \texttt{NullPointerException} & Null access during \texttt{PUBLISH} message handling. \\ 
12 & MQTT   & Moquette  & \texttt{StacklessClosedChannelException} & Connection closed before sending \texttt{CONNACK}. \\ \hline
\end{tabular}
    
\end{table*}

\begin{table*}[tbp]
\centering

\caption{The comparison of test-case generation between \ourmethod{} and \fume{} over three runs (on Mosquitto).}
\label{tab:rq2_generation}

\begin{tabular}{c|ccc|ccc}
\toprule
\multirow{2}{*}{\textbf{Run}} & \multicolumn{3}{c|}{\textbf{\ourmethod{}}} & \multicolumn{3}{c}{\textbf{\fume{}}} \\
\cmidrule{2-7}
 & Total Cases & Unique Cases & Avg. Length & Total Cases & Unique Cases & Avg. Length \\
\midrule
1 & \textbf{3,333,073} & \textbf{1,796,307} & 127.06 & 1,875,234 & 1,138,751 & \textbf{139.26} \\
2 & \textbf{2,125,908} & \textbf{1,179,660} & \textbf{137.91} & 1,874,732 & 1,137,339 & 135.46 \\
3 & \textbf{1,610,051} & 899,816   & \textbf{144.34} & 1,519,627 & \textbf{925,739}   & 129.8 \\
\bottomrule
\end{tabular}

\end{table*}

\section{Evaluation and Discussion}

By evaluating across general-purpose protocols and comparing with both traditional and LLM-based fuzzers, we demonstrate that \ourmethod{} is broadly applicable, capable of uncovering real-world bugs, and cost-efficient in terms of token usage.

\subsection{RQ1: Flexibility and Effectiveness Across Protocol Implementations}

\ourmethod{} successfully discovered multiple bugs across diverse protocol implementations, demonstrating both effectiveness and flexibility. Table~\ref{tab:bug_summary} summarizes the 12 potential bugs discovered by \ourmethod{} across multiple protocol implementations.

For the MQTT protocol, we evaluated three broker implementations: HMQ, Moquette, and Mosquitto. On HMQ, we identified a critical bug that causes the system to crash. On Moquette, \ourmethod{} uncovered three distinct issues that trigger exceptions, though the broker remains operational. No bugs were discovered in Mosquitto, likely because we reused the payload generator from \fume{}, which has already been used extensively to test Mosquitto, and most known issues have been patched.

Listing \ref{lst:bugmq} shows the buggy code we discovered in the HMQ MQTT broker. The issue lies in the conditional statement {\lstset{basicstyle=\ttfamily\normalsize,language=Go}\lstinline|conn != nil && conn.RemoteAddr() != nil|}. While it appears safe, calling {\lstset{basicstyle=\ttfamily\normalsize,language=Go}\lstinline!conn.RemoteAddr()!} when {\lstset{basicstyle=\ttfamily\normalsize,language=Go}\lstinline!conn!} is {\lstset{basicstyle=\ttfamily\normalsize,language=Go}\lstinline!nil!} will still cause a runtime panic in Go, as method calls on a {\lstset{basicstyle=\ttfamily\normalsize,language=Go}\lstinline!nil!} interface result in dereferencing a {\lstset{basicstyle=\ttfamily\normalsize,language=Go}\lstinline!nil!} pointer. This leads to a crash if the code path is ever executed with a {\lstset{basicstyle=\ttfamily\normalsize,language=Go}\lstinline!nil!} {\lstset{basicstyle=\ttfamily\normalsize,language=Go}\lstinline!conn!}, which our generated input successfully triggered. This example illustrates how \ourmethod{} can uncover subtle but critical edge-case bugs by exploring under-tested execution paths.

\begin{lstlisting}[label=lst:bugmq, language=Go, caption={Buggy code in HMQ where line 2 contains a faulty conditional check: calling \lstinline!conn.RemoteAddr()! without ensuring \lstinline!conn! is non-\lstinline!nil! leads to a runtime panic.}]
// add remote connection address
if !wsEnabled && conn != nil && conn.RemoteAddr() != nil 
{
    result = append(result, zap.Stringer("addr", conn.RemoteAddr()))
} 
else if wsEnabled && wsConn != nil && wsConn.Request() != nil 
{
    result = append(result, zap.String("addr", wsConn.Request().RemoteAddr))
}
\end{lstlisting}

For Modbus, we tested the Pymodbus implementation. Our approach identified eight unique bugs, each causing exceptions without crashing the system. They are all related to incorrect buffer lengths during binary unpacking. These exceptions point to improper handling of specific malformed or edge-case inputs and reflect the method’s ability to exercise error-handling paths even in well-established protocol libraries.

\begin{lstlisting}[label=lst:bugmd, language=Python, caption={Buggy code in Pymodbus where line 1 attempts to unpack 5 bytes from a buffer without validating its length, leading to a \texttt{struct.error} when the data is too short.}]
self.address, count, _byte_count = struct.unpack(">HHB", data[0:5])
\end{lstlisting}

The bug occurs in the Modbus implementation when parsing a request frame with an unexpected or malformed length. Specifically, the code in Listing~\ref{lst:bugmd} attempts to unpack 5 bytes from the incoming \lstset{basicstyle=\ttfamily\normalsize,language=Python}\lstinline!data! buffer using \lstset{basicstyle=\ttfamily\normalsize,language=Python}\lstinline!struct.unpack(">HHB", data[0:5])!. However, if the \lstset{basicstyle=\ttfamily\normalsize,language=Python}\lstinline!data! is shorter than 5 bytes, it results in a \lstset{basicstyle=\ttfamily\normalsize,language=Python}\lstinline!struct.error! with the message \texttt{unpack} requires a buffer of 5 bytes. This indicates a missing length check before unpacking, which can cause the program to crash or throw an exception at runtime.

In the case of DAAP, we tested the OwnTone media server. Our testing did not uncover new bugs. One challenge here is that DAAP is layered over HTTP, and our system currently models only the DAAP-specific parts, without generating complete HTTP requests. Furthermore, OwnTone has already been tested by \chatafl{}~\cite{meng2024large}, which may have addressed some common issues. Nevertheless, \ourmethod{} was still able to process and explore DAAP's binary structure with minimal manual adjustment, underlining its general applicability.

\subsection{RQ2: Effectiveness vs. Model-Based Fuzzers}

To evaluate how our LLM-assisted fuzzing approach compares to existing model-based fuzzers, we conducted a head-to-head comparison with \fume{}, a domain-specific MQTT fuzzer. Our evaluation consists of two parts: the first measures the ability of each tool to generate diverse and voluminous test cases within a fixed time budget; the second assesses the efficiency of each method in triggering faults by comparing the time taken to induce a crash.

Table~\ref{tab:rq2_generation} presents the test case generation results of our approach and \fume{} across three independent one-hour fuzzing runs on the Mosquitto MQTT broker. \ourmethod{} consistently generates a higher number of total and unique test cases compared to \fume{}. For example, in the first run, \ourmethod{} produced over 3.3 million total cases and nearly 1.8 million unique ones, whereas \fume{} generated only 1.8 million total cases and 1.1 million unique cases. This trend holds across all three runs, demonstrating the higher test case diversity and generation throughput of our LLM-assisted approach.

While the average payload length varies slightly, both tools produce messages of comparable size, indicating similar levels of complexity in the generated data. This suggests that \ourmethod{} is not simply generating larger or noisier inputs to inflate test coverage, but is instead producing diverse, well-formed messages that are competitive in structure and semantics. Moreover, the consistently higher number of unique cases indicates broader exploration of the input space, which can lead to uncovering more edge cases and rare protocol behaviors.

\begin{figure}[btp]
\centering
\includegraphics[width=0.45\textwidth]{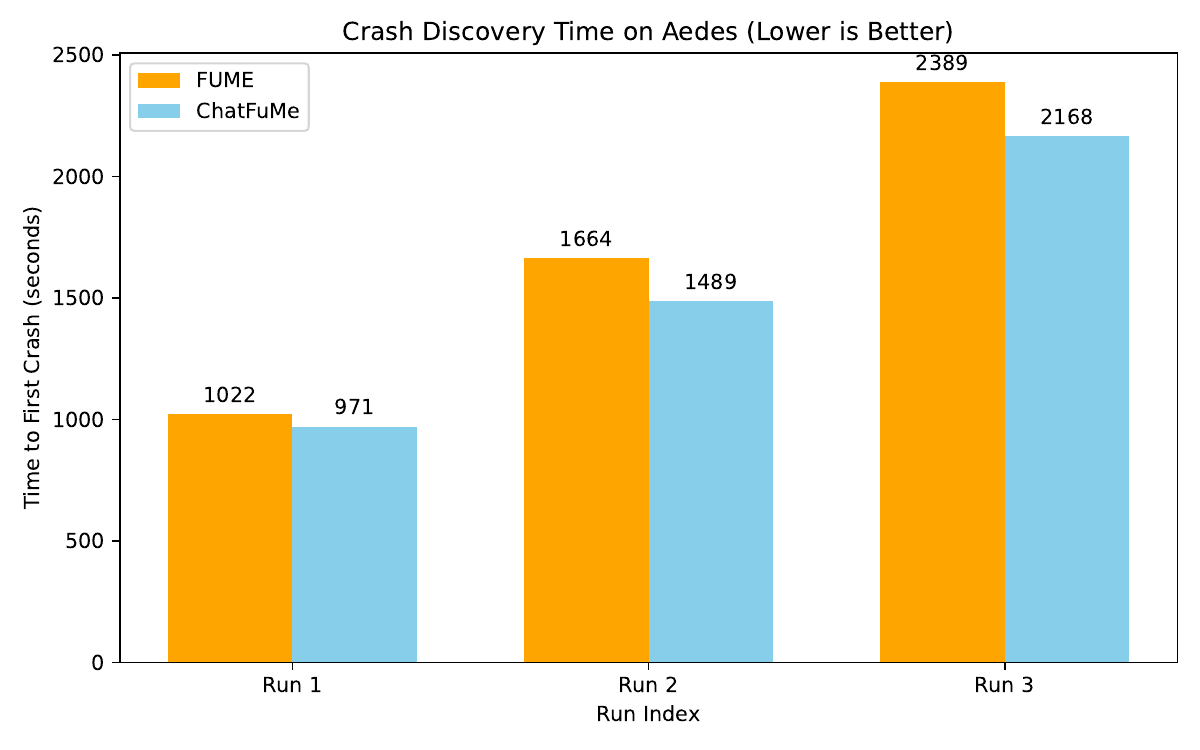}
\caption{The comparison of crash discovery time between \ourmethod{} and \fume{} over three runs (on Aedes).}
\label{fig:crash_comparison}
\end{figure}

To assess crash discovery efficiency, we compared our LLM-assisted fuzzer against \fume{} on the Aedes MQTT broker under identical conditions over three trials. \ourmethod{} located the first crash in an average of 1543.0 seconds, compared to 1691.7 seconds for \fume{}, which is an improvement of roughly 9\%. In every individual run, our approach detected faults faster (16m11s vs. 17m02s, 24m50s vs. 27m44s, and 36m08s vs. 39m49s), demonstrating that our LLM-guided sequence generation can accelerate the identification of critical vulnerabilities.

\subsection{RQ3: Cost Efficiency vs. LLM-Based Fuzzers}

To assess the efficiency of \ourmethod{} compared to existing LLM-based fuzzing approaches, we conducted a one-hour experiment using both \ourmethod{} and \chatafl{}.
We measured the total number of tokens consumed and the number of LLM calls made during the fuzzing process.
As shown in Table~\ref{tab:rq3_tokens}, our approach consumed only 5,980 tokens and made 16 LLM calls, while \chatafl{} consumed 216,596 tokens and issued 160 LLM calls in the same period.
This demonstrates that \ourmethod{} is more token-efficient---using roughly 36 times fewer tokens and 10 times fewer LLM calls---making it more practical for long-running or cost-sensitive fuzzing campaigns.

While the token and call statistics provide a clear comparison of efficiency, it is important to note that a direct comparison of the generated test cases between \ourmethod{} and \chatafl{} is limited due to differences in how the two tools store their outputs. Specifically, \chatafl{} only retains test cases that trigger new execution paths, whereas \ourmethod{} stores all generated cases for analysis and replay. Upon inspection of the test cases retained by \chatafl{}, we found that they primarily consisted of HTTP-like requests such as \texttt{GET} and \texttt{POST}, which aligns with the nature of DAAP as an HTTP-based protocol. In contrast, the test cases produced by \ourmethod{} are raw hexadecimal payloads that conform to the DAAP message format.

\begin{table}[btp]
\centering

\caption{The comparison of LLM token and API call usage between \ourmethod{} and \chatafl{} over a one-hour run.}
\label{tab:rq3_tokens}

\begin{tabular}{l|c|c}
\hline
\textbf{Method} & \textbf{Tokens Used} & \textbf{LLM Calls} \\
\hline
\ourmethod{} & 5,980  & 16  \\
\chatafl{}    & 216,596 & 160 \\
\hline
\end{tabular}

\end{table}

\subsection{Discussion}

\paragraph{Limitations}
We do not include an ablation study in this work because our goal is to minimize manual effort and demonstrate how LLMs can be leveraged to model protocol structures with minimal human input automatically. Rather than manually deconstructing and varying components, our focus is on showcasing the feasibility and effectiveness of using LLMs in a streamlined and integrated way. The strength of our approach lies in its simplicity and automation. We focus our evaluation on comparing the complete system against established baselines, including traditional fuzzers and existing LLM-based methods.

Our findings suggest that LLMs can be used not only to automate the creation of fuzzers but also to generate malicious tools for exploiting vulnerabilities. This highlights a broader concern about the dual-use nature of large language models, suggesting that future work should consider safeguards and responsible deployment practices.

\paragraph{Theat to construct validity}
One limitation of our method, common to many fuzzing techniques, is that it primarily exposes surface-level bugs: crashes or exceptions caused by malformed inputs. While our method generates syntactically diverse and realistic messages, it does not capture deeper semantic behaviors that may be necessary to uncover subtle logic bugs. This may limit the types of vulnerabilities our tool can detect, potentially underestimating deeper security issues present in the target systems.

\paragraph{Threat to internal validity}
Malformed inputs may not directly cause some bugs discovered during fuzzing, but rather be caused by unrelated factors such as system configuration or dependency issues. We mitigate this by confirming that crashes are triggered deterministically with repeated inputs. 

\paragraph{Threat to external validity}
Our evaluation focuses on a select set of protocols (MQTT, Modbus, DAAP) and open-source implementations. While they cover multiple transport layers and application domains, generalizing our results to all protocol-based software may be limited. Proprietary systems, real-time protocols, or those with more complex state machines may exhibit different behavior.

In some cases, the initial protocol specifications used by our payload generator were derived from LLM output (e.g., ChatGPT). While this allows automation, it also introduces potential inaccuracies or omissions compared to official standards. The effectiveness of our fuzzing may partially rely on the correctness of LLM-generated specifications, which may not generalize well to protocols with complex or poorly documented semantics.

\section{Related Work}
\label{sec:related}

\subsection{Protocol Implementation Fuzzing}

Protocol implementation fuzzing is a testing technique that systematically sends malformed and unexpected inputs to network protocol implementations to uncover bugs, vulnerabilities, or unexpected behavior. Fuzzing techniques in terms of input generation are commonly categorized into generation-based and mutation-based approaches. Generation-based fuzzing~\cite{beurdouche2017messy, reen2020dpifuzz, somorovsky2016systematic, fiterau2023automata} constructs inputs from predefined specifications, ensuring syntactic correctness and compliance with the protocol. Mutation-based fuzzing~\cite{amusuo2023systematically, he2022intelligent, pham2020aflnet, natella2022stateafl} modifies existing valid inputs to create test cases, relying on randomness or heuristics to explore unexpected behaviors. A common approach for fuzzers to improve performance on semantic constraints is to build a protocol communication model. The model enables fuzzers to generate structured and context-sensitive message sequences. \fume{}~\cite{pearson2022fume} manually constructs a communication model for MQTT and integrates generation-based and mutation-based fuzzing in this model. There are some works that use automated methods to build communication models~\cite{luo2019polar, zhao2019seqfuzzer, gascon2015pulsar}. For example, Pulsar~\cite{gascon2015pulsar} automatically builds a communication model by analyzing traffic loads. 

There are also several LLM-based fuzzers designed for testing protocol implementations. These approaches typically provide the LLM with protocol inputs or documentation and prompt it to generate test cases in the form of protocol payloads. \chatafl{}~\cite{meng2024large} is a general fuzzing framework that directly uses LLM to extract information and generate initial inputs. In this framework, LLM plays an important role in initializing the seed of fuzzing and provides guidance for mutation based on coverage. However, \chatafl{} is primarily designed for string-based protocol implementations, leveraging the strengths of LLMs in understanding and generating structured text data. mGPTFuzz~\cite{ma2024one} is an LLM-based fuzzing framework for Matter IoT Devices~\cite{matterIOt}. In mGPTFuzz's fuzzing loop, LLM is first asked to extract information from Matter’s specification. Users then prompt the LLM to build finite state machines (FSMs) based on the extracted information. Finally, it generates inputs based on FSMs and a user-defined policy. LLMIF~\cite{ wang2024llmif} is an LLM-based fuzzing framework for Zigbee IoT devices~\cite{gislason2008zigbee}, utilizing LLM in the process of protocol information extraction and response reasoning.

Our work differs from existing approaches in several ways. Compared to manually crafted model-based fuzzers and protocol-specific LLM-based fuzzes like mGPTFuzz, our method uses LLMs to automatically build and adapt different protocol models, making it more general and less dependent on expert input. In contrast to LLM-based fuzzers that directly generate individual test cases, our method uses the LLM to program a protocol-aware sequence generator, providing better control over input structure and reducing token consumption.

\subsection{LLM in Testing} 
Large language models (LLMs) have shown strong performance across multiple tasks in software engineering~\cite{schafer2023empirical, li2024enhancing, kang2024quantitative}. By leveraging their ability to understand and generate code, LLMs can help identify edge cases, create meaningful test inputs, and detect potential vulnerabilities. Fuzz4all\cite{xia2024fuzz4all} is a universal fuzzing framework for compiler testing. It outperforms different baseline tools in 6 different programming languages. Whitefox~\cite{yang2024whitefox} is a white-box fuzzer for testing logic bugs. SymPrompt\cite{ryan2024code} presents a prompting strategy for test generation. By implementing a multi-stage workflow, SymPrompt reaches higher code coverage in several open-source Python projects. 

Our work differs from existing LLM-based fuzzers in other domains, such as compiler testing or code-based test generation, by focusing on general protocol implementation fuzzing. Unlike those approaches that typically generate code or API calls as test cases, our method finally generates protocol message payloads that must conform to specific communication sequences. This requires handling stateful interactions and semantic constraints unique to network protocols, which we address by using LLMs to model protocol behavior and guide input generation, rather than producing test cases directly.

\subsection{Feedback-Guided Testing}

Feedback-guided testing \cite{cai2002optimal} is a software cybernetics \cite{cai2003overview} approach to software testing where test strategies evolve dynamically based on real-time feedback from previous test executions. Unlike traditional static testing methods, this approach treats the software under test as a controlled object and the testing process as a feedback control loop. The central idea is to collect the outcome data from the executed test cases. For instance, in the Controlled Markov Chain model \cite{hu2008adaptive}, the testing process is governed by an estimated state, and optimal actions are selected to meet reliability goals with minimal resource consumption.

Our method incorporates the idea of feedback-guided testing by using LLMs to iteratively adjust the input generator based on feedback from previous test executions. We adapt the generator by modifying protocol states or transitions, allowing the fuzzer to explore diverse and previously untested behaviors. This approach brings the principles of ART into the LLM era, enabling automated, feedback-driven refinement of test strategies in a structured and scalable way.
\section{Conclusion}
In this work, we propose a novel LLM-assisted fuzzing method \ourmethod{} that automates protocol modeling and test case generation with minimal manual effort. \ourmethod{} demonstrates strong flexibility across multiple protocols and is effective in identifying real-world bugs in diverse software systems. Through extensive evaluation, we show that our approach generates more diverse and higher-volume test cases than traditional fuzzers, while being significantly more efficient in discovering crashes. Moreover, our method achieves these results using far fewer LLM tokens and calls compared to other LLM-based fuzzers, highlighting its efficiency and practicality. This study illustrates the potential of large language models to streamline and enhance fuzz testing, opening new directions for intelligent and automated software testing.

\bibliographystyle{ACM-Reference-Format}
\bibliography{sample-base}

\end{document}